# Scale analysis for the Madden–Julian oscillation

Hing Ong[a]

[a] *University of California, Davis, Davis, CA, USA*

*Corresponding author*: Hing Ong, hing5ong5@gmail.com

ABSTRACT

This study performs a scale analysis for the dynamics of equatorial flow systems like the Madden–Julian oscillation (MJO). This analysis makes three assumptions about the geometry of the flow and focuses on the leading order terms of the mass continuity and momentum equation system. First, the flow is meridionally narrow. Second, the zonal structure is planetary wavenumber 1. Last, the flow is vertically shallow. When constrained by these geometric assumptions, the system shows a time scale much longer than a day depending on the meridional narrowness. Specifically, for equatorial Kelvin waves, the oscillation period ranges from 36 to 64 days given a ratio of the planetary radius to the meridional width scale from 6 to 8 (i.e., an $e$-folding latitude from 9.59° to 7.18° either side of the equator). Given the long, narrow, and shallow geometry, the dynamics can sustain Kelvin waves only at low frequencies. This result suggests that the intraseasonal time scale of the MJO arises from the Kelvin wave frequency limit due to the planetary wave geometry. This planetary intraseasonal scale relation is independent from thermodynamics, and an open question remains on what physical processes coordinate the thermodynamics in the same length or time scale.

## 1. Introduction

The intraseasonal period (30 to 60 days) of the Madden–Julian oscillation (MJO) is a mystery of great importance (Jiang et al., 2020; Zhang, 2005; Zhang et al., 2020). Xie et al. (1963) first documented the tropical intraseasonal oscillation of winds (Li et al., 2018; Xie et al., 2018). Then, Madden and Julian (1972) discovered the planetary-scale alternation of zonal winds, pressure anomalies, and convectively active-inactive phases propagating eastward along the equator later called the MJO. Later studies revealed that the MJO modulates high-impact weather and climate phenomena not only in the tropics (e.g., Xie et al., 1963) but also in the middle to high latitudes through teleconnections (Jiang et al., 2020; Zhang, 2013). Also, the MJO serves as a primary source of global system predictability in the subseasonal time scale (Neena et al., 2014; Waliser, 2012). The structure and evolution of the MJO is well observed (Jiang et al., 2020). For example, contours of MJO's geopotential anomalies zonally span about $180^{\circ}$ but meridionally extend only about $10^{\circ}$ either side of the equator (Adames and Wallace, 2014; Leung and Qian, 2017). Many theories have been developed for the MJO, and most of them agree that convective heating plays a crucial role (Jiang et al., 2020; Zhang et al., 2020). However, the fundamental physics behind the planetary length scale and intraseasonal time scale have remained uncertain as those theories vastly disagree on this matter (Jiang et al., 2020; Zhang et al., 2020).

The development of MJO theory can be traced back to Chang (1977) and other early studies (e.g., Chao, 1987; Emanuel, 1987; Lau and Peng, 1987; Wang, 1988; Wang and Chen, 1989). This effort was further motivated by Wheeler and Kiladis (1999), who identified the convectively coupled equatorial waves (see also Kiladis et al., 2009). They smoothed the power spectrum of tropical cloud brightness and found several spectral peaks standing out from the smoothed spectrum in the wavenumber-frequency space. They explained the distrubution of all the peaks using the theoretical equatorial wave solutions derived by Matsuno (1966) except a peak of planetary-scale intraseasonal eastward-propagating signals associated with the MJO. Specifically, Wheeler and Kiladis (1999) matched the observed spectral distrubution with Matsuno's (1966) solutions by tuning the theoretical shallow water equivalent depth. Since Matsuno's (1966) theory only missed the MJO under a certain range of the equivalent depth, the development of MJO theory became a major target of scientific research. Notably, the Kelvin wave band follows the theory in wavenumber 2 or higher. However, at wavenumber 1, the theory expects a 20-day period by extrapolation, but the MJO dominates with longer periods.

Matsuno's (1966) theory neglects many physical processes, and all MJO theories aim to identify the physical processes essential to the MJO. Despite the agreement on the importance of convective heating, different closure assumptions for the convective heating have been used (Zhang et al., 2020). Moreover, each MJO theory adds different prognostic variables or closure assumptions. Many MJO theories need a prognostic moisture variable (Adames and Kim, 2016; Ahmed, 2021; Fuchs and Raymond, 2017; Majda and Stechmann, 2009; Raymond, 2001; Wang et al., 2016). Among them, some require cloud-radiative heating closure (Adames and Kim, 2016; Wang et al., 2016), and Majda and Stechmann (2009) introduced the unique wave activities as a prognostic variable. These theories stand for an idea that moisture field evolution is essential to the MJO. However, some other theories explain the MJO while neglecting moisture field evolution. For example, Yang and Ingersoll (2013) used convective trigger instead of moisture to close the convective heating. Also, Yano and Tribbia (2017) and Rostami and Zeitlin (2020) considered nonlinear potential vorticity advection as the primary propagation mechanism. In addition, in pursuit of a simple explanation for the intraseasonal time scale, Kim and Zhang (2021) assumed linear momentum damping, for which the mechanism is yet to be identified. To summarize, the essential physical processes of the MJO remain as a matter of debate.

The purpose of this study is to investigate whether the intraseasonal time scale of the MJO is due to the planetary length scale. Instead of adding physical assumptions to build on Matsuno's (1966) theory, this study makes geometric scaling assumptions to the governing equations and seeks a relation between the planetary length scale and intraseasonal time scale. This approach stems from the mathematical overdetermination of Matsuno's (1966) solutions for Kelvin waves, which propagate eastward along the equator. There are three equations:

$$\frac{\partial \varphi}{\partial t} + H_{\mathrm{E}} \frac{\partial u}{\partial x} = 0, \tag{0a}$$

$$\frac{2\Omega}{a} y u + \frac{\partial \varphi}{\partial y} = 0, \tag{0b}$$

$$\frac{\mathrm{D}u}{\mathrm{D}t} + \frac{\partial \varphi}{\partial x} = 0. \tag{0c}$$

However, there are only two unknown variables: $u$, zonal velocity; and $\varphi$, pressure or geopotential (height or pressure coordinates). The parameters are defined as follows: $H_{\mathrm{E}}$, equivalent depth; $\Omega$, planetary rotation rate; $a$, planetary radius. The conventional solutions for Kelvin waves are based on Equations (0a) and (0c), and Equation (0a) is based on thermodynamic assumptions (e.g., Vallis, 2017, Kiladis et al., 2009). On the other hand, this study makes geometric assumptions and solves for Kelvin waves using Equations (0b) and (0c). Here Kelvin waves refer to waves characterized by pressure gradient in the zonal direction and geostrophic balance in the meridional direction. Given this definition, Kelvin waves are not necessarily gravity waves since Equation (0a) is not needed. For the rest of this paper, Section 2 presents the scale analysis, and Section 3 summarizes and discusses the results.

## 2. Scale Analysis

Several previous studies have performed scale analyses for the MJO and other tropical atmospheric flows (Yano and Bonazzola, 2009, hereafter YB09; cf. Adames, 2022; Adames et al., 2019; Gill, 1980; Heckley and Gill, 1984; Ogrosky and Stechmann, 2015). They found geostrophic balance in the meridional but not zonal direction when the meridional-zonal aspect ratio is small (Ogrosky and Stechmann, 2015; see also Adames, 2022; Adames et al., 2019; Gill, 1980; Heckley and Gill, 1984). Also, they related most characteristic time scales to thermodynamics. As a notable exception, YB09 found that the planetary wave time scale only depends on the length scale. However, YB09 applied the same length scale to meridional and zonal directions. The present analysis follows YB09's approach but considers a small aspect ratio. In short, this analysis first derives a pressure scale from meridional geostrophic balance and then a time scale of zonal wind responding to this pressure scale. Detailed analysis starts from the mass continuity, meridional momentum, and zonal momentum equations:

$$\frac{\partial u}{\partial x}+\frac{\partial v}{\partial y}+\frac{\partial w}{\partial z}=0, \tag{1a}$$

$$\frac{\mathrm{D}v}{\mathrm{D}t}+\frac{2\Omega}{a}yu+\frac{\partial \varphi}{\partial y}=0, \tag{1b}$$

$$\frac{\mathrm{D}u}{\mathrm{D}t}-\frac{2\Omega}{a}yv+2\Omega w+\frac{\partial \varphi}{\partial x}=0. \tag{1c}$$

The variables $v$ and $w$ denote meridional and vertical velocity. The operator $\frac{\partial}{\partial z}$ denotes density-weighted vertical divergence. The shallow water approximation has been relaxed because it is not needed. There are 4 unknown variables but only 3 equations. More equations are needed to derive flow solutions except Kelvin wave solutions, where $\hat{v}=0$ (Matsuno, 1966). From here, this analysis makes three scaling assumptions.

The first scaling assumption is meridional narrowness. Specifically, in Equations (1b) and (1c), the traditional and nontraditional Coriolis parameters are linearized as $\frac{2\Omega}{a}y$ and $2\Omega$ (e.g.,

Ong and Roundy 2019; 2020). This approximation is used in Matsuno (1966) and most MJO theories (Zhang et al., 2020) and is well justified by the meridional narrowness of the observed geopotential anomalies of the MJO (Adames and Wallace, 2014; Leung and Qian, 2017).

The second scaling assumption is a planetary zonal length scale. The selection of the planetary length scale is a target of many MJO theories (Zhang et al., 2020). In other words, they make closure assumptions to find the physics behind the length scale. Nevertheless, the purpose of this analysis is to find a time scale given a length scale. Hence, a dimensionless zonal length is defined as $\hat{x} = \frac{sx}{a}$, where $s$ denotes planetary wavenumber. Accordingly, Equations (1) are scaled with $\hat{x}$ along with other dimensionless variables defined as follows: $\hat{y} \equiv \frac{y}{Y}$, $\hat{z} \equiv \frac{z}{H}$, $\hat{t} \equiv Ft$, $\hat{u} \equiv \frac{u}{U}$, $\hat{v} \equiv \frac{v}{V}$, $\hat{w} \equiv \frac{w}{W}$, and $\hat{\varphi} \equiv \frac{\varphi}{\Phi}$. Plugging them into Equations (1) and rearranging yield the followings:

$$\frac{sYU}{aV}\frac{\partial \hat{u}}{\partial \hat{x}} + \frac{\partial \hat{v}}{\partial \hat{y}} + \frac{YW}{HV}\frac{\partial \hat{w}}{\partial \hat{z}} = 0, \tag{2a}$$

$$\frac{FaV}{2\Omega YU}\frac{\mathrm{D}\hat{v}}{\mathrm{D}\hat{t}} + \hat{y}\hat{u} + \frac{a\Phi}{2\Omega Y^2 U}\frac{\partial \hat{\varphi}}{\partial \hat{y}} = 0, \tag{2b}$$

$$\frac{FaU}{2\Omega YV}\frac{\mathrm{D}\hat{u}}{\mathrm{D}\hat{t}} - \hat{y}\hat{v} + \frac{aW}{YV}\hat{w} + \frac{s\Phi}{2\Omega YV}\frac{\partial \hat{\varphi}}{\partial \hat{x}} = 0. \tag{2c}$$

The third scaling assumption is vertical shallowness. Specifically, the flow is shallow enough for the nontraditional Coriolis term $\frac{aW}{YV}\hat{w}$ to be negligible (e.g., Ong and Roundy 2019; 2020). Matsuno (1966) and most MJO theories (Jiang et al., 2020; Zhang et al., 2020) used this approximation, and this study provides a justification. Following Ong and Roundy (2019; 2020), the nondimensional parameter for the nontraditional Coriolis term is $\frac{aW}{YV} \cong \frac{aH}{Y^2}$. Given $a$ = 6,380 km, $H \cong$ 16 km, and $Y \sim$ 1000 km for the MJO, the $\frac{aH}{Y^2}$ parameter is 10%. Hereafter, the nontraditional Coriolis term drops out of the equations.

The rest of this analysis first considers zonal divergence on the leading order and then relaxes this assumption with a meridional Rossby number.

*a. Zonal Divergence Scaling*

Zonal divergence is significant for the MJO and should scale on the leading order of Equation (2a). Hence, $V = \frac{sUY}{a}$ is required, and Equations (2) become as follows:

$$\frac{\partial \hat{u}}{\partial \hat{x}} + \frac{\partial \hat{v}}{\partial \hat{y}} + \frac{YW}{HV}\frac{\partial \hat{w}}{\partial \hat{z}} = 0, \tag{3a}$$

$$\frac{sF}{2\Omega}\frac{\mathrm{D}\hat{v}}{\mathrm{D}\hat{t}} + \hat{y}\hat{u} + \frac{a\Phi}{2\Omega Y^2 U}\frac{\partial\hat{\varphi}}{\partial\hat{y}} = 0, \tag{3b}$$

$$\frac{Fa^2}{2\Omega sY^2}\frac{\mathrm{D}\hat{u}}{\mathrm{D}\hat{t}} - \hat{y}\hat{v} + \frac{a\Phi}{2\Omega Y^2 U}\frac{\partial\hat{\varphi}}{\partial\hat{x}} = 0. \tag{3c}$$

The pressure gradient must be nonzero for nontrivial solutions, and quasi-geostrophic zonal winds are necessary to sustain nonzero pressure gradient in Equation (3b) ($\Phi = \frac{2\Omega Y^2 U}{a}$). As for Equation (3c), the zonal wind tendency should respond to the zonal pressure gradient in a frequency scale much lower than the planetary rotation rate ($F \ll 2\Omega$) depending on the meridional narrowness ($Y \ll a$):

$$F = 2\Omega s\frac{Y^2}{a^2}. \tag{4}$$ [1]

Equation (4) is based on the following deductive reasoning. If $F$ was higher, the $\Phi/U$ ratio would be higher and result in imbalance in Equation (3b). If $F$ was lower, an additional term in Equation (3c) would be needed to balance the pressure gradient along the equator ($\hat{y} = 0$). For example, if zonal momentum damping was added to balance the pressure gradient, the pressure gradient would not necessarily drive a zonal wind tendency, allowing a lower $F$. Similarly, if the flow was vertically deep enough to retain the nontraditional Coriolis term, vertical motion could reach quasi-geostrophic balance, allowing a lower $F$. If the meridional-zonal aspect ratio was unity ($\frac{sY}{a} = 1$), Equation (4) would reduce to $F = 2\Omega\frac{Y}{a}$, consistent with YB09. For the meridionally narrow and vertically shallow case, Equation (4) holds, and consequently, Equations (3) become as follows:

$$\frac{\partial\hat{u}}{\partial\hat{x}} + \frac{\partial\hat{v}}{\partial\hat{y}} + \frac{YW}{HV}\frac{\partial\hat{w}}{\partial\hat{z}} = 0, \tag{5a}$$

[1] MacDonald's (2024) equation (5) implies this relation between the time scale and the length scale assuming $v = 0$. This analysis explains this relation in detail without that assumption. I thank Kai-Chih Tseng for bringing MacDonald (2004) to my attention after publication of the present paper.

$$\frac{s^2Y^2}{a^2}\frac{\mathrm{D}\hat{v}}{\mathrm{D}\hat{t}} + \hat{y}\hat{u} + \frac{\partial\hat{\varphi}}{\partial\hat{y}} = 0, \tag{5b}$$

$$\frac{\mathrm{D}\hat{u}}{\mathrm{D}\hat{t}} - \hat{y}\hat{v} + \frac{\partial\hat{\varphi}}{\partial\hat{x}} = 0. \tag{5c}$$

Kelvin wave solutions emerge from linearized Equations (5b) and (5c) as follows:

$$\hat{u} = \hat{\varphi} = \exp\left[-\frac{y^2}{\left(\sqrt{2}Y\right)^2} + i\left(\frac{sx}{a} - Ft\right)\right]. \tag{6}$$

The wave amplitude decays with increasing distance from the equator with an $e$-folding length of $\sqrt{2}Y$. Also, according to Equation (4), these waves oscillate with the following period ($T$):

$$T \equiv \frac{2\pi}{F} = \frac{a^2}{s\left(\sqrt{2}Y\right)^2}\frac{2\pi}{\Omega} \cong \frac{a^2}{s\left(\sqrt{2}Y\right)^2}\text{day} = \frac{1}{s\sin^2\vartheta}\text{day}, \tag{7}$$

where $\vartheta$ denotes the $e$-folding latitude. Equation (7) suggests that the oscillation period of these waves is a day multiplied by the square of the ratio of planetary radius to the $e$-folding length given $s = 1$. Given the observed spatial structure of the MJO (e.g., Adames and Wallace, 2014; Leung and Qian, 2017), this ratio roughly ranges from 6 to 8 ($e$-folding latitude from 9.59° to 7.18° either side of the equator). Then, this scale analysis predicts an oscillation period ranging from 36 to 64 days, which matches the intraseasonal time scale of the MJO. This analysis also results in a phase speed ($Fa$) from 12.92 to 7.27 m s$^{-1}$ eastward, which is about twice the MJO's observed interquartile phase speed range of 3 to 6 m s$^{-1}$ (Chen and Wang, 2020). These results suggest that the intraseasonal time scale and eastward phase speed of the MJO can be explained to the leading order by the Kelvin wave dynamics constrained by the long, narrow, and shallow geometry. The phase speed difference may be due to other physical processes, plausibly momentum damping (Chang, 1977; Kim and Zhang, 2021) or background zonal wind advection (Roundy, 2022; Roundy and De Castro, 2024).

The linear approximation requres limited scales of wind speed and geopotential. The material tendency operator $\frac{\mathrm{D}}{\mathrm{D}t}$ comprises a linear local tendency and three nonlinear advection terms: $F\frac{\partial}{\partial\hat{t}} + \frac{sU}{a}\hat{u}\frac{\partial}{\partial\hat{x}} + \frac{V}{Y}\hat{v}\frac{\partial}{\partial\hat{y}} + \frac{W}{H}\hat{w}\frac{\partial}{\partial\hat{z}}$. The advection scale ($\frac{sU}{a}$) should not exceed $F$, and the geopotential scale ($\Phi$) should not exceed geostrophic scaling. Accordingly:

$$U \leq 2\Omega\frac{Y^2}{a}, \tag{8a}$$

$$V \leq 2\Omega s\frac{Y^3}{a^2}, \tag{8b}$$

$$\Phi \leq \frac{4\Omega^2 Y^4}{a^2}. \tag{8c}$$

For the linear MJO scaling given $s = 1$, the theoretical upper bounds of zonal speed, meridional speed, and geopotential height anomalies are 12.92 m s$^{-1}$, 2.15 m s$^{-1}$, and 17.0 m, which are comparable to observations (Kiladis et al., 2005; Leung and Qian, 2017; Madden and Julian, 1972)

As a side note, low-wavenumber equatorial Rossby waves can be derived from Equations (5) along with closure assumptions (e.g., Gill, 1980; Heckley and Gill, 1984; Vallis, 2017). Equatorial Rossby waves are important in tropical intraseasonal oscillation theories. For example, the westward intraseasonal oscillation can be explained as convectively coupled equatorial Rossby waves (Chen, 2022; Fuchs-Stone et al., 2019; Kiladis et al., 2009; Wheeler and Kiladis, 1999). Also, the MJO are associated with Rossby wave structures in observations, and some theories consider that Rossby wave dynamics are important to the MJO (Ahmed, 2021; Rostami and Zeitlin, 2020; Yano and Tribbia, 2017). The present study complements those studies with an alternative explanation for the intraseasonal time scale from a geometric perspective.

*b. Meridional Rossby Number Scaling*

Given Equation (2b), the meridional Rossby number ($Ro$) is defined as the ratio of the meridional wind tendency to the meridional Coriolis term:

$$Ro \equiv \frac{FaV}{2\Omega YU}. \tag{9}$$

This definition uses the tendency time scale instead of the advective time scale used in the conventional definition (e.g., Vallis, 2017). The derivation of Equations (5a) and (5b) suggests that the zonal divergence scales on the leading order only if $Ro \approx \frac{s^2 Y^2}{a^2}$. To relax this assumption, this analysis substitutes Equation (9) into Equations (2) and considers geostrophic scaling in Equation (2b) ($\frac{s^2 Y^2}{a^2} \leq Ro \leq 1$):

$$\frac{1}{Ro}\frac{sF}{2\Omega}\frac{\partial \hat{u}}{\partial \hat{x}} + \frac{\partial \hat{v}}{\partial \hat{y}} + \frac{YW}{HV}\frac{\partial \hat{w}}{\partial \hat{z}} = 0, \tag{10a}$$

$$Ro\frac{\mathrm{D}\hat{v}}{\mathrm{D}\hat{t}} + \hat{y}\hat{u} + \frac{\partial \hat{\varphi}}{\partial \hat{y}} = 0, \tag{10b}$$

$$\frac{1}{Ro}\frac{F^2 a^2}{4\Omega^2 Y^2}\frac{\mathrm{D}\hat{u}}{\mathrm{D}\hat{t}} - \hat{y}\hat{v} + \frac{1}{Ro}\frac{sF}{2\Omega}\frac{\partial \hat{\varphi}}{\partial \hat{x}} = 0. \tag{10c}$$

For the zonal wind tendency to remain on the leading order of Equation (10c), $Ro$ should be proportional to the square of the frequency scale:

$$F = \sqrt{Ro}\frac{2\Omega Y}{a}. \tag{11}$$

Accordingly, Equations (10) become as follows:

$$\frac{1}{\sqrt{Ro}}\frac{sY}{a}\frac{\partial\hat{u}}{\partial\hat{x}} + \frac{\partial\hat{v}}{\partial\hat{y}} + \frac{YW}{HV}\frac{\partial\hat{w}}{\partial\hat{z}} = 0, \tag{12a}$$

$$Ro\frac{\mathrm{D}\hat{v}}{\mathrm{D}\hat{t}} + \hat{y}\hat{u} + \frac{\partial\hat{\varphi}}{\partial\hat{y}} = 0, \tag{12b}$$

$$\frac{\mathrm{D}\hat{u}}{\mathrm{D}\hat{t}} - \hat{y}\hat{v} + \frac{1}{\sqrt{Ro}}\frac{sY}{a}\frac{\partial\hat{\varphi}}{\partial\hat{x}} = 0. \tag{12c}$$

Section 2a discusses the flow regime of $Ro = \frac{s^2Y^2}{a^2}$, and here, changing $Ro$ leads to other flow regimes given the aspect ratio $\frac{sY}{a}$ while the hatted variables still scale as unity. According to Equations (11), (12a), and (12c), increasing $Ro$ from $\frac{s^2Y^2}{a^2}$ allows a higher frequency $F$ at the expense of losing planetary-scale pressure gradient and mass divergence in the zonal direction, which are inversely proportional to the square root of $Ro$. This scale separation is vast at wavenumber 1 ($s = 1$), making Kelvin waves unsustainable with higher frequencies.

Taking the limit of $Ro = 1$, wavenumber-0 mixed Rossby-gravity mode solutions (Matsuno, 1966) emerge from the leading order of linearized Equations (12) as follows:

$$\hat{v} = \exp\left[-\frac{y^2}{\left(\sqrt{2}Y\right)^2} + iFt\right], \tag{13a}$$

$$\hat{u} = \hat{\varphi} = -i\frac{y}{Y}\exp\left[-\frac{y^2}{\left(\sqrt{2}Y\right)^2} + iFt\right]. \tag{13b}$$

The oscillation amplitude maxima are located at a distance $Y$ from the equator. Also, Equation (11) yields the following period ($T$):

$$T \equiv \frac{2\pi}{F} = \frac{2\pi a}{2\Omega Y} = \frac{a}{2Y}\frac{2\pi}{\Omega} \cong \frac{a}{2Y}\,\mathrm{day}, \tag{14}$$

Given an $a$-to-$Y$ ratio from 6 to 8, the period $T$ ranges from 3 to 4 days, which matches the observed low-wavenumber convectively coupled mixed Rossby-gravity mode (Wheeler and Kiladis, 1999). The other high-frequency Matsuno's (1966) modes, namely the inertio-gravity modes, can be found by repeating Equations (13) with higher Hermite polynomials.

This analysis suggests that the Rossby number poses an important limit to Kelvin wave frequency at wavenumber 1 according to Equations (11) and (12). With a low Rossby number (oscillation period of 36 to 64 days), the pressure gradient and mass divergence necessary for Kelvin waves are strong enough (Figure 1a). As Rossby number increases, frequency becomes higher, but Kelvin waves cease because of the lack of pressure gradient and mass divergence in the zonal direction. Meanwhile, the pressure gradient and mass divergence in the meridional direction are still available for the mixed Rossby-gravity (Figure 1b) and inertio-gravity wave modes. As for the equatorial Rossby wave modes, the Rossby number is intrinsically low. Thus, the frequency limit due to the planetary wave geometry is unique to the Kelvin wave mode while the characteristics of all other wave modes should still be explained by the theoretical shallow water equivalent depth (Matsuno, 1966).

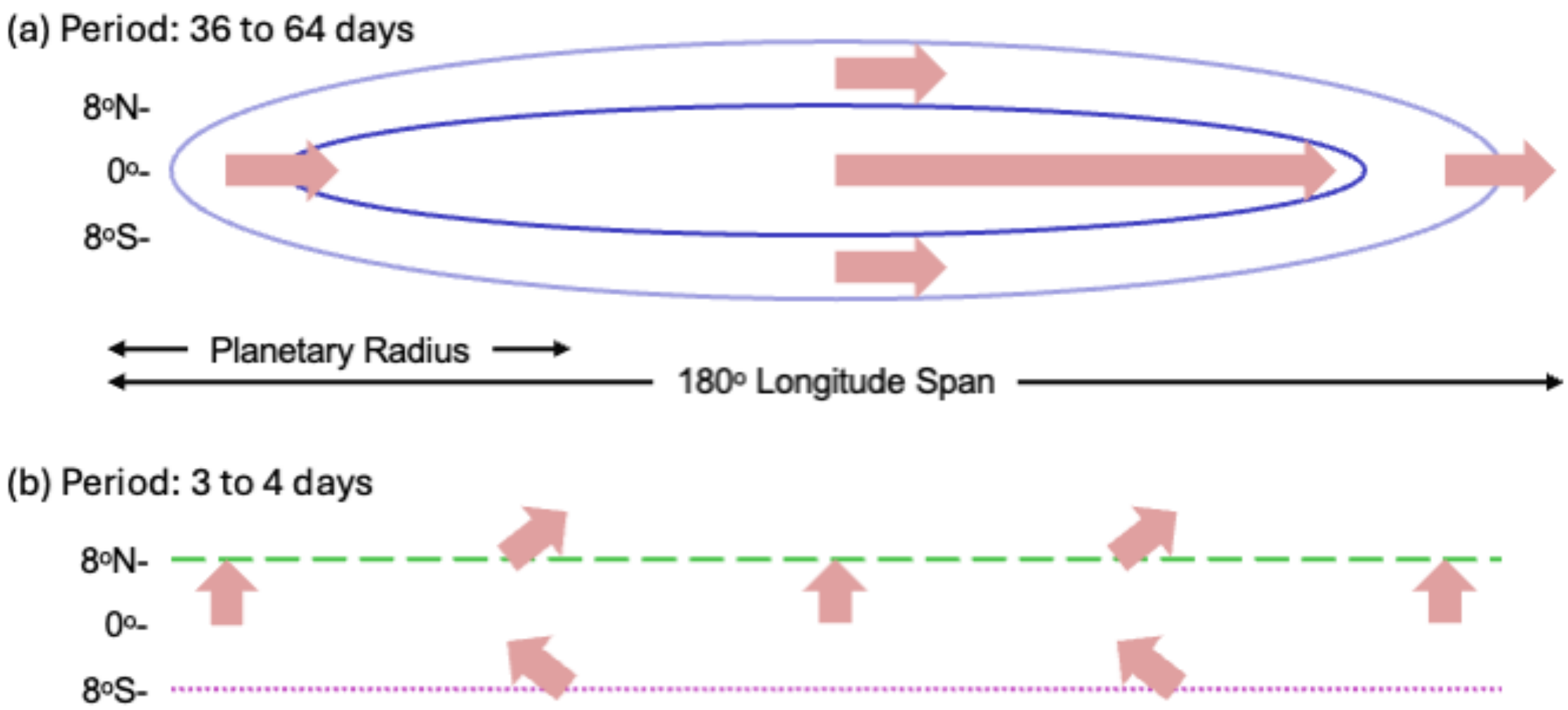

Figure 1. Characteristic snapshots of equatorial planetary-scale fluid dynamics. Panel (a) depicts a low Rossby number example with a period of 36 to 64 days, featuring pressure contours and wind arrows of the Kelvin wave mode in Equation (6). Panel (b) depicts a high Rossby number example with a period of 3 to 4 days, featuring lines of pressure ridge (dashed) and trough (dotted) along with wind arrows of the mixed Rossby-gravity wave mode in Equations (13). The contours denote $e^{-4}$ and $e^{-1}$ of the pressure maximum.

Finally, for wavenumber 2 or higher ($s \geq 2$), the meridional-zonal aspect ratio $\frac{sY}{a}$ becomes larger, and the scale of zonal mass divergence as well as pressure gradient is no longer vastly separated from the leading order. Thus, Kelvin waves are sustainable in a broad frequency range (i.e., a wide range of $Ro$) for wavenumber 2 or higher, which explains not only the spectral band marked as Kelvin waves in Wheeler and Kiladis (1999) but also the observed Kelvin wave signals in lower frequencies (Roundy, 2012; 2020). However, Kelvin waves are limited to frequencies associated with the MJO (i.e., low $Ro$ only) at wavenumber 1. The extension of the MJO band into higher wavenumbers may be due to other physical processes. For example, the slowness of the MJO in the Eastern Hemisphere causes the MJO signals to project onto wavenumber 2 (Roundy, 2012; 2022; Roundy & De Castro, 2024).

## 3. Summaries and Discussions

Diverse MJO theories have been developed to fill the gap between the theory by Matsuno (1966) and the observations by Wheeler and Kiladis (1999). Each MJO theory adds different physical processes, leaving no satisfactory conclusion on what is the essential process. This study focuses on the geometric constraints on Kelvin wave dynamics without adding any physical process. The results suggest that Matsuno's (1966) Kelvin waves of wavenumber 1 oscillate with an intrinsic period of 36 to 64 days, which matches the observed MJO period. This conclusion is based on three assumptions. First, the flow is meridionally narrow. Second, the zonal structure is planetary wavenumber 1. Last, the flow is vertically shallow. All these assumptions are well-justified.

Moving toward higher frequency along wavenumber 1, Kelvin waves become dynamically unsustainable. This unique frequency limit due to the planetary length scale does not affect other wave modes found by Matsuno (1966). Also, moving toward higher wavenumbers, this frequency limit becomes less restrictive on Kelvin wave dynamics. Therefore, the theoretical shallow water equivalent depth should remain in control on the characteristics of equatorial waves (Matsuno 1966) except wavenumber-1 Kelvin waves, which is controlled by the planetary wave geometry.

Arguments taking Kelvin waves as the core of the MJO dynamics can be traced back to Chang (1977) and are still used in recent studies (Kim and Zhang, 2021; Roundy, 2022; Roundy and De Castro, 2024). This study attributes the intraseasonal time scale to the long, narrow, and shallow geometry, which complements their arguments with a simple explanation. The present scale analysis results are independent from thermodynamics, which is consistent with Yano and Bonazzola (2009). The remaining question is the closure of the equation set that coordinates the thermodynamic equation in the same length or time scale, where convective heating should play a crucial role.

*Acknowledgments.*

This study began as Ryan Torn assigned a homework question in February 2018 in the Synoptic Dynamic Meteorology course at University at Albany, State University of New

York, stating "Derive the QG momentum equations for a location near (not at) the equator, where $L_x$ = 10,000 km, but $L_y$ = 1000 km." I brought this question to University of California, Davis and kept studying it for recreation. Discussions with Paul Roundy, Da Yang, Joseph Biello, Seth Seidel, Argel Ramírez Reyes, and Lin Yao helped building the ideas. I completed this study during the winter break in December 2024. Comments from the editor and three anonymous reviewers improved the presentation of this paper.

*Data Availability Statement.*

No datasets were generated or analyzed during the current study.